# Assistance orale à la recherche visuelle

*Etude expérimentale de l'apport d'indications spatiales à la détection de cibles*

*Speech Assistance to Visual Search: Experimental Assessment of the Utility and Usability of Spatial Indications for Visual Target Detection*

**KIEFFER Suzanne, CARBONELL Noëlle**

LORIA, France
Suzanne.Kieffer@loria.fr, Noelle.Carbonell@loria.fr

**Résumé.** La parole associée au geste, une forme courante d'expression multimodale destinée à l'utilisateur, a fait l'objet de nombreuses études ergonomiques. En revanche, rares sont les recherches publiées sur la conception et l'évaluation d'interfaces où les réactions du système combinent présentations graphiques et messages oraux. Cet article décrit une étude expérimentale visant à évaluer la contribution effective de messages oraux à l'efficacité et au confort d'activités de recherche visuelle. La tâche choisie est le repérage et la sélection à la souris, dans une collection de 30 photographies affichées à l'écran, de la seule photographie familière aux participants car elle leur a été présentée auparavant. 24 participants ont effectué 240 tâches de repérage visuel dans deux conditions se distinguant uniquement par la présentation initiale de la cible : visuelle, PV, ou multimodale, PM ; PM comprenait une présentation visuelle de la cible accompagnée d'un message oral d'aide à sa localisation ultérieure dans la collection affichée. Les participants avaient pour consigne d'effectuer le repérage aussi rapidement que possible. Les temps moyens de sélection des cibles se sont avérés trois fois plus longs dans la condition PV que dans la condition PM, et les erreurs presque deux fois plus nombreuses. En outre, les performances des participants et l'efficacité des messages oraux ont été influencées par l'organisation spatiale de l'affichage des collections et la difficulté des tâches visuelles. En particulier, les messages oraux se sont avérés particulièrement utiles pour les tâches visuelles difficiles. La plupart de ces résultats sont statistiquement significatifs. En outre, cette forme d'assistance orale au repérage de cibles a été bien acceptée par l'ensemble des sujets dont la majorité a préféré la présentation multimodale de la cible à sa présentation exclusivement visuelle.

**Mots-clés.** assistance orale à la recherche visuelle, détection visuelle de cibles, étude expérimentale, évaluation ergonomique, interaction homme-machine multimodale, organisation spatiale des affichages 2D, parole et graphique.

**Abstract.** The utility and usability of speech as a supplementary input modality have been extensively investigated. Contrastingly, issues relating to the implementation of speech synthesis as an additional expression modality for the system have motivated few ergonomic research studies until now. We report an experimental study aiming at assessing the possible contribution of oral system messages to the efficiency and comfort of visual search, an activity which users perform frequently while interacting with standard graphical user interfaces, for instance, whenever they manipulate icons or browse photos. Each experimental task consisted in detecting and selecting (using the mouse) a previewed photo in a collection of 30 colour photos pulled from popular Web sites and displayed on a 21" screen. 24 participants performed 240 such visual search tasks in two conditions (120 tasks per condition) which only differed from each other in the initial presentation of the target photo: visual (VP) versus multimodal (MP). In the MP condition, the presentation included, in addition to the target visual presentation, oral cues on its location in the collection of photos displayed next. Participants were instructed to perform target search and selection as fast as they could. Average selection times proved to be three times longer in the VP condition than in the MP one, and detection error rates were twice higher. In addition, oral messages were well accepted by most participants who preferred multimodal target presentations to visual ones. Besides, results demonstrate the influence of task difficulty and collection display layout on participants' performances and message efficiency. Oral messages proved to be most useful for difficult tasks, and visual search performances were lower for currently used 2D array layouts than for two out of the three other spatial structures tested.
**Keywords.** oral assistance to visual search, visual target detection, experimental study, usability study, multimodal human-computer interaction, spatial layout of 2D displays, speech and graphics

## 1 Introduction

Les avancées scientifiques et technologiques récentes permettent de diversifier les médias et modalités d'interaction Homme-Machine, et de créer de nouvelles formes de multimodalité. L'utilisateur peut désormais interagir avec le système par la parole, grâce aux progrès de la reconnaissance vocale, par le geste manuel avec un doigt, un stylo ou un gant numérique (Ehrenmann *et al.*, 2001), ou encore par le regard, notamment dans le domaine médical (Blois *et al.*, 1999). En sortie, les réactions de la machine aux commandes ou actions des utilisateurs sont plus limitées en raison principalement de la diversité moindre des dispositifs de sortie.

La multimodalité qui associe, en entrée, la parole à d'autres modalités a suscité de nombreux travaux de recherche. En revanche, l'association, en sortie, de la parole au graphique et/ou au texte n'a motivé que quelques études. On peut citer à titre d'exemple, celle de Faraday et Sutcliffe (1997) sur l'évaluation de documents multimédias constitués de graphiques accompagnés de commentaires oraux. Les possibilités offertes par la parole en tant que modalité d'expression du système complémentaire du texte et/ou du graphique ont été encore moins explorées, à notre connaissance. Alors même que les problèmes d'ordre logiciel sont résolus depuis longtemps, l'étude de la contribution de la parole à l'efficacité des interventions du système (messages d'erreurs, comptes rendus d'exécution, aide en ligne) en est encore à ses débuts. Or, on ne peut raisonnablement proposer à l'utilisateur une interface avec laquelle il pourrait interagir oralement mais qui resterait muette.

Dans ce contexte, nous avons choisi de nous intéresser à la multimodalité parole plus graphique en tant que mode d'expression du système. Plus précisément, nous avons étudié l'apport de messages système oraux à l'efficacité et au confort d'activités de recherche visuelle. Le terme « graphique » désigne ici et dans toute la suite, outre les objets graphiques classiques (schémas, icônes, …), les images statiques, telles que dessins, peintures, photographies, notamment lorsqu'il est associé à « modalité » ou « affichage ».

Cet article présente une étude expérimentale portant sur l'évaluation ergonomique de la contribution d'informations spatiales orales à la précision, la rapidité et la facilité du repérage visuel de cibles au sein d'affichages 2D denses et complexes de collections de photographies. Dans le contexte de cette étude, le terme « précision » fait référence au succès ou à l'échec du repérage de la cible. Nous présentons d'abord le contexte scientifique, les motivations et les objectifs de notre recherche. Nous décrivons ensuite notre démarche, la méthodologie adoptée et le protocole expérimental mis en œuvre, avant de présenter et discuter les résultats expérimentaux obtenus.

## 2  Contexte scientifique, motivations et objectifs

### 2.1  Vers de nouvelles formes d'interaction Homme-Machine

Étudier, en situation d'interaction Homme-Machine, l'association de la parole au graphique dans une même intervention du système est un sujet de recherche qui présente actuellement un intérêt particulier.

D'une part, l'intégration de la parole aux modalités de sortie actuelles est susceptible d'enrichir l'interaction en mobilisant deux facultés sensorielles de l'utilisateur, l'audition et la vision, au lieu d'une seule, la vision. En outre, si la manipulation directe, mode d'interaction prévalant aujourd'hui, suffit pour interagir avec les applications classiques qui combinent graphique et texte en sortie, la diversification des contextes d'utilisation fait de la parole une modalité d'interaction utile, voire indispensable dans certaines situations.

D'autre part, les concepteurs d'applications multimédias incluent dans les présentations visuelles un nombre croissant d'éléments sonores tels que messages vocaux, bips ou musique, sans s'interroger sur les effets éventuels de cet enrichissement des médias de présentation sur la charge cognitive de l'utilisateur. Associer parole et graphique dans une même intervention du système soulève des problèmes ergonomiques qui n'ont pas encore été abordés, sauf pour des catégories spécifiques d'utilisateurs (Yu et Brewster, 2003) ou pour des contextes spécifiques d'utilisation (Wang *et al.*, 2000). Des recherches dans ce domaine sont donc nécessaires pour être en mesure de proposer aux concepteurs des recommandations ergonomiques qui permettent une intégration de la parole aux modalités classiques d'expression du système satisfaisante pour l'utilisateur.

### 2.2  Surcharge croissante des affichages et navigation

Parallèlement, les progrès des techniques d'affichage ont entraîné le développement rapide de nouvelles fonctions d'interaction, telles que vue d'ensemble, zoom, filtrage, ou encore visualisation des relations entre les objets graphiques affichés (Shneiderman, 1996). En outre, le flot des informations visuelles transmises à l'utilisateur est en constante augmentation (Krause, 1997). Ces évolutions se traduisent par une surcharge des présentations visuelles qui ne cesse de croître. La multiplication des fenêtres, des barres d'outils et des icones affichées simultanément, ainsi que le développement des mises en relief visuelles et des animations, sont la source, pour l'utilisateur, de difficultés d'interaction supplémentaires, perceptives (perception de l'environnement graphique) et cognitives (augmentation de la charge de travail). Cette augmentation rapide de la densité des affichages graphiques est de nature à ralentir et rendre fastidieuse et fatigante la recherche visuelle (Pirolli, 2000). En effet, on a montré que la densité des informations affichées affecte l'attention visuelle en réduisant la taille du champ visuel utile (U.F.O.V. pour Useful Field Of View). En outre, la recherche dans un affichage est moins efficace si celui-ci comprend des zones plus denses que d'autres (Drury et Clement, 1978).

Pour pallier ces problèmes et offrir à l'utilisateur un accès visuel rapide à des données de types variés, de nombreuses techniques de visualisation d'informations ont été élaborées. On peut citer, à titre d'exemple, les visualisations scientifiques (graphes, diagrammes, graphes multidimensionnels, etc.), les diagrammes de noeuds et de liens, les Tree-Maps (Card et Mackinlay, 1997). Ces techniques, qui s'appuient sur la cartographie et la sémiologie graphique (Bertin, 1983), visent à assister l'utilisateur dans l'analyse de volumes de données importants en lui fournissant, sous la forme d'indices visuels faciles à appréhender, des informations sur les propriétés de ces données et les relations qui existent entre elles (Ware, 2004).

Plus généralement, la recherche et la navigation dans les grands ensembles d'informations exigent un effort cognitif important de la part des utilisateurs, même avec l'assistance de techniques récentes de visualisation multi-échelle préservant le contexte : vues en œil de poisson (Furnas, 1986), murs en perspective (Mackinlay *et al.*, 1991), arbres hyperboliques (Lamping *et al.*, 1995), superposition de vues transparentes (Harisson et Vicente, 1996).

### 2.3 Motivations et objectifs

Nous avons choisi d'étudier et de clarifier les apports potentiels de la parole en tant que modalité de sortie complémentaire du graphique, dans un contexte où le rôle des messages oraux est d'assister l'utilisateur dans une activité de recherche visuelle. Le domaine d'application choisi est la recherche d'informations dans les banques d'images. Il s'agit d'une application en plein essor, touchant à la fois le grand public et les spécialistes. En outre, la recherche visuelle dans ce contexte est une activité complexe et exigeante ; assister l'utilisateur dans cette activité est donc utile, et lui fournir une telle assistance sous forme orale plutôt que visuelle afin de ménager sa vision est justifié.

L'évaluation ergonomique de cette forme d'assistance à l'exploration d'affichages complexes est un sujet de recherche qui, à notre connaissance, n'a pas encore été abordé dans un contexte d'interaction homme-machine, bien que l'exploration de scènes compte, avec la lecture, parmi les activités visuelles qui ont suscité le plus de travaux de recherche en psychologie, mais uniquement dans le contexte de tâches expérimentales en laboratoire ; voir par exemple (Findlay et Gilchrist, 1998) et (Henderson et Hollingworth, 1998). En outre, la parole est considérée, à juste titre, comme le mode le plus naturel de communication humaine, en particulier pour ce qui est des échanges d'informations. L'intégration de cette modalité aux interfaces Homme-Machine, qu'elles soient destinées au grand public ou aux professionnels, doit donc permettre d'accroître sensiblement la diffusion de l'informatique dans la société.

En bref, notre ambition est de proposer une forme d'intégration de la parole aux modalités de sortie actuelles qui mette à profit l'enrichissement de l'interaction résultant de cette intégration pour assister l'utilisateur dans ses activités de recherche visuelle sans faire appel à ses capacités de vision de plus en plus sollicitées.

Les applications potentielles d'une telle étude ne se réduisent pas au domaine de la visualisation interactive de grands ensembles d'informations. L'informatique mobile ou embarquée a entraîné également une augmentation sensible des informations visuelles qu'il est nécessaire de transmettre à l'utilisateur, par exemple sur son itinéraire (e.g., cartes, plans). Il en est de même pour les services Web multimédias dont l'exploration est rendue de plus en plus difficile en raison de l'augmentation des icones, dessins, images, photographies, etc.

## 3   Démarche

L'étude expérimentale présentée vise à déterminer l'influence d'indications spatiales orales destinées à faciliter la recherche d'informations dans des affichages complexes, sur les performances et la satisfaction d'utilisateurs potentiels.

### 3.1   Choix de l'activité

La situation d'interaction multimodale retenue est le repérage de photographies connues visuellement, désigné dans la suite par les expressions « repérage visuel » ou « détection de cibles ». Les messages oraux, très brefs pour ne pas créer de surcharge cognitive, situent la position de la cible par rapport à neuf zones pré-définies sur l'écran.

Nous avons retenu la détection de cibles comme tâche expérimentale car elle intervient dans de nombreuses activités interactives :

- L'exploration de visualisations de grands ensembles d'informations (données scientifiques, résultats d'une requête en fouille de données) ou la navigation sur Internet ; dans ces situations, un système adaptatif peut déterminer les informations susceptibles d'intéresser l'utilisateur et les localiser oralement dans l'affichage courant.
- L'inspection visuelle assistée par ordinateur, au cours de la fabrication de produits industriels ou bien pour la maintenance ou la sécurité (Drury, 1992) ; le système, lorsqu'il détecte une anomalie peut la signaler oralement à l'utilisateur en la localisant dans l'affichage courant ; on peut généraliser cette forme d'assistance aux applications de contrôle de processus impliquant une activité d'inspection ou de surveillance visuelle exigeante, par exemple, le contrôle du trafic aérien ou des centrales nucléaires, le pilotage d'avions.
- L'interaction avec des systèmes iconiques ou des applications graphiques aux affichages denses et complexes, qui lie directement la détection de cibles visuelles à l'action (e.g., clics souris, actions de cliquer-glisser, etc.) en créant une « boucle sensori-motrice » (Rasmussen, 1986) ; l'inclusion, dans les messages d'aide en ligne, d'informations sur la position des icônes et des objets graphiques évoqués dans ces messages peut faciliter au novice leur repérage dans l'affichage ; voir, concernant l'apport de la parole aux modalités classiques d'expression des informations d'aide, (Capobianco et Carbonell, 2003).

Bien qu'il existe en psychologie de nombreux travaux de recherche publiés sur le repérage de cibles, la plupart d'entre eux portent sur des tâches artificielles de laboratoire. Voir, par exemple, les travaux de Kramer *et al.* (2001) sur l'évolution des performances de détection de cibles en présence de distracteurs, en fonction de leur nombre et de leurs propriétés visuelles. On peut citer également ceux de Diederich *et al.* (2003) sur l'incidence des interférences entre perception visuelle et tactile sur les temps de réaction des sujets. D'autres disciplines comme les neurosciences se sont également intéressées à la recherche visuelle, mais dans une perspective différente de la nôtre, l'exploration des mécanismes neuronaux sous-jacents à cette activité chez les humains et les singes (Chelazzi 1999). Enfin, en ergonomie des logiciels, l'assistance orale à la recherche visuelle dans des situations d'interaction classiques n'a encore fait l'objet d'aucune étude à notre connaissance.

### 3.2 Choix des informations orales destinées à l'utilisateur

Nous avons choisi d'étudier l'influence d'informations orales spatiales concernant la position de la cible dans l'affichage, sur l'efficacité et le confort de son repérage. On s'attend, en réduisant la zone de recherche, à une réduction du temps de détection. Il existe en français deux catégories d'informations spatiales qui, en situation de dialogue oral, peuvent constituer des énoncés complets : les indications absolues, qui expriment la position d'un élément dans l'affichage, et les indications relatives, qui le localisent par rapport à un autre élément affiché. Par exemple, l'énoncé « En haut. » lorsqu'il est synonyme de « En haut de l'écran/de la fenêtre. » appartient à la première catégorie, tandis que « A gauche de l'icône de justification. » appartient à la seconde. A noter que ces expressions font référence implicitement au point de vue de l'utilisateur : gauche et droite sont définies par rapport à lui. Nous avons retenu les indications spatiales absolues comme forme d'assistance à la localisation des cibles, car elles sont plus simples que les indications relatives, donc plus rapides à interpréter et moins sujettes à l'ambiguïté.

On peut donc préciser comme suit la question de recherche qui motive l'étude présentée. Des messages sonores contenant des informations spatiales absolues sur la position de la cible dans l'affichage sont-ils susceptibles d'améliorer significativement la précision et la rapidité de sa détection ? La réponse à cette question peut paraître évidente. Cependant, les résultats d'une première expérimentation présentée dans (Carbonell et Kieffer, 2002 et 2005) ont montré que des indications orales sur la position de la cible dans l'image étaient sans effet sur les temps de détection. Quant à l'influence éventuelle de ce type d'information sur la précision du repérage, le protocole expérimental adopté ne permettait pas de l'étudier. Sur le plan théorique, la situation choisie pose une question restée sans réponse satisfaisante jusqu'à présent, en raison des connaissances limitées dont on dispose sur les rôles respectifs des processus intentionnels et perceptifs (i.e., contrôlés par les stimuli) dans la commande des mouvements oculaires et le contrôle de l'attention visuelle, ainsi que sur leurs interactions éventuelles. Au cas particulier, quels sont les effets, sur les mouvements oculaires des utilisateurs, donc sur leurs performances et leur confort visuel, de l'influence exercée simultanément sur ces mouvements par la saillance visuelle des constituants des affichages d'une part, et par les indications spatiales associées à ces affichages d'autre part ? Des interférences entre processus cognitifs et perceptifs de contrôle de l'attention visuelle pourraient expliquer l'absence d'influence sensible des indications spatiales sur les temps de détection des cibles observés lors de notre première étude. D'un point de vue pratique, orienté vers la conception d'applications nouvelles, il est nécessaire d'obtenir des éléments de réponse précis à cette question théorique, avant d'envisager la réalisation d'aides en ligne orales à la recherche visuelle et leur intégration aux applications qui engagent l'utilisateur dans des activités exigeantes de recherche visuelle. En effet, il ne suffit pas qu'une telle assistance ait un effet positif sur l'efficacité et le confort de ces activités, encore faut-il que son apport soit suffisamment important pour que les utilisateurs l'acceptent et l'utilisent régulièrement, ne la considèrent pas comme un « gadget » superflu, et donc pour que les concepteurs envisagent son intégration effective à des applications réelles.

### 3.3 Choix des situations expérimentales étudiées

Pour évaluer l'apport effectif des messages oraux définis en 3.2 au repérage de cibles, nous avons comparé les performances d'utilisateurs potentiels dans deux situations d'interaction Homme-Machine :

- la recherche d'un objet/élément graphique connu visuellement dont la position dans l'affichage est inconnue ;
- la recherche d'un objet/élément graphique connu visuellement dont la position dans l'affichage est indiquée de façon approximative par l'un des messages suivants : « En haut/bas. », « A droite/gauche. » ou « Au centre. ».

La réalisation de chaque situation impose, pour que l'utilisateur puisse se familiariser visuellement avec la cible, de lui présenter d'abord la cible isolée, puis d'afficher la scène, dans laquelle il doit la rechercher. Chaque tâche de recherche visuelle comprend donc deux étapes : la première, consacrée à la présentation de la cible, la seconde, comprenant l'affichage de la scène, la recherche et la sélection de la cible. Les conditions expérimentales associées à ces deux situations ne se distinguent l'une de l'autre que par le mode de présentation de la cible :

- visuel : affichage (sur un écran de 21") de la cible isolée pendant 3 secondes ; le format d'affichage de la cible, dans la présentation et dans la scène, est le même ;
- multimodal : affichage de la cible isolée, accompagné simultanément d'un message oral donnant des informations sur sa position dans la scène (durée de la présentation : 3 secondes également).

Pour un couple (cible, affichage) donné, la seconde étape consiste, dans les deux conditions, à rechercher et sélectionner la cible dans la scène affichée. Pour que la position de la souris soit identique pour tous les sujets et toutes les tâches, les sujets doivent déclencher l'affichage de la scène en cliquant sur un bouton au centre de l'écran. Le clic de sélection dans la scène déclenche le passage à la tâche suivante. Dans le cadre de notre étude, la présentation exclusivement visuelle des cibles correspond à la condition de référence.

## 4 Méthodologie

### 4.1 Hypothèses de travail

L'expérience présentée vise à tester l'hypothèse suivante : la présentation multimodale de la cible améliore-t-elle de façon significative, par rapport à sa présentation visuelle, la précision et la rapidité de son repérage dans l'image. (H1)

D'autre part, nous avons montré dans notre première étude (Carbonell et Kieffer, 2002 et 2005), sur une tâche peu différente, l'influence de la structure visuelle des scènes sur les performances des utilisateurs : la moitié des échecs observés lors de la sélection à la souris des cibles étaient dus à l'absence d'organisation spatiale de l'affichage. Ces premiers résultats nous ont conduits à tenir compte, dans la présente étude, de l'influence possible de l'organisation spatiale des affichages sur le repérage de cibles, avec ou sans l'assistance d'indications spatiales. Cette expérience vise donc également à tester l'hypothèse suivante : la structure visuelle des images ou scènes présentées à l'utilisateur est-elle susceptible d'influer sur ses performances dans les deux situations. (H2)

Tester ces deux hypothèses présente un double intérêt. Sur le plan scientifique, d'une part, car elles n'ont pas encore fait l'objet d'évaluations systématiques, en particulier dans des situations réalistes d'interaction Homme-Machine. Sur le plan des perspectives d'applications, d'autre part, car les organisations spatiales adoptées pour la visualisation d'ensembles de données scientifiques, d'icônes ou d'images n'ont fait l'objet d'aucune évaluation ergonomique spécifique, à notre connaissance.

### 4.2 Caractéristiques des affichages

**_Affichages 2D statiques_**

Les techniques d'affichage en trois dimensions ont connu récemment un développement spectaculaire : il est désormais possible d'interagir avec le système au sein d'espaces virtuels 3D. Il convient de noter cependant que l'interaction avec les interfaces graphiques 3D, en l'état actuel, engendre pour l'utilisateur de nouvelles difficultés. Par exemple, la profusion des couleurs ne facilite pas la perception de l'espace. De même, l'évaluation des positions ou des mouvements relatifs des objets, la planification des chemins d'accès à l'information sont des tâches difficiles qu'il serait utile d'assister (Ware, 2004). Par ailleurs, si les affichages 3D suscitent généralement des jugements subjectifs positifs, l'efficacité (rapidité et précision) des interactions qu'ils permettent est inférieure à celle qu'autorisent les affichages interactifs 2D (Levy *et al.*, 1996 ; Sutcliffe et Patel, 1996). C'est pour ces raisons que nous avons choisi de limiter notre champ d'étude aux représentations en deux dimensions. Nous avons également restreint, dans un premier temps, la portée de notre recherche à la détection de cibles dans des affichages statiques. En effet, la détection de cibles dans des présentations animées est une activité visuelle différente, bien plus complexe, qui demande des recherches spécifiques.

**_Structures aléatoire, elliptique, matricielle et radiale_**

Quatre organisations spatiales des affichages ont été testées dans le cadre de cette étude (voir la figure 1) :
- aléatoire, où les éléments de la scène sont placés aléatoirement ;
- elliptique, où ils sont disposés le long de deux ellipses concentriques ;
- matricielle, où ils forment un tableau à deux dimensions ;
- radiale, où ils sont disposés en étoile (8 branches).

Le choix de structures symétriques simples, matricielle, radiale et elliptique, se justifie par le fait que ces trois structures se rapprochent d'organisations souvent utilisées pour afficher des informations. On peut considérer les structures radiale et elliptique comme des représentations schématisées des arbres hyperboliques générés par l'une des principales techniques de visualisation d'ensembles structurés d'informations structurés (Lamping *et al.*, 1995). La forme de la structure elliptique s'apparente également à celle des cadrans circulaires. Quant à la structure matricielle, elle reproduit l'organisation spatiale utilisée le plus souvent pour présenter les documents contenus dans un dossier sous forme d'icônes, et les photographies d'une collection sous forme de vignettes. Enfin, la détection visuelle d'une cible dans un affichage non structuré (i.e., organisé aléatoirement) peut être assimilée à la recherche d'une icône, sur le bureau d'un ordinateur par exemple.

### 4.3 Caractérisation du matériel visuel

Afin de rendre la tâche de repérage la plus réaliste possible, nous avons centré l'expérience sur la recherche de photographies représentant des objets et des paysages réels. La tâche expérimentale consiste donc à rechercher une photographie particulière dans une collection de photographies affichées suivant l'une des quatre structures retenues, puis à la sélectionner à la souris le plus rapidement possible. Elle est réaliste car elle est proche des recherches visuelles que l'on est amené à effectuer lors de la consultation d'une banque d'images. Dans la suite de la présentation, nous utiliserons le terme « collection » pour désigner les photographies qui composent un affichage, et nous appellerons « scène » le contenu d'un affichage.

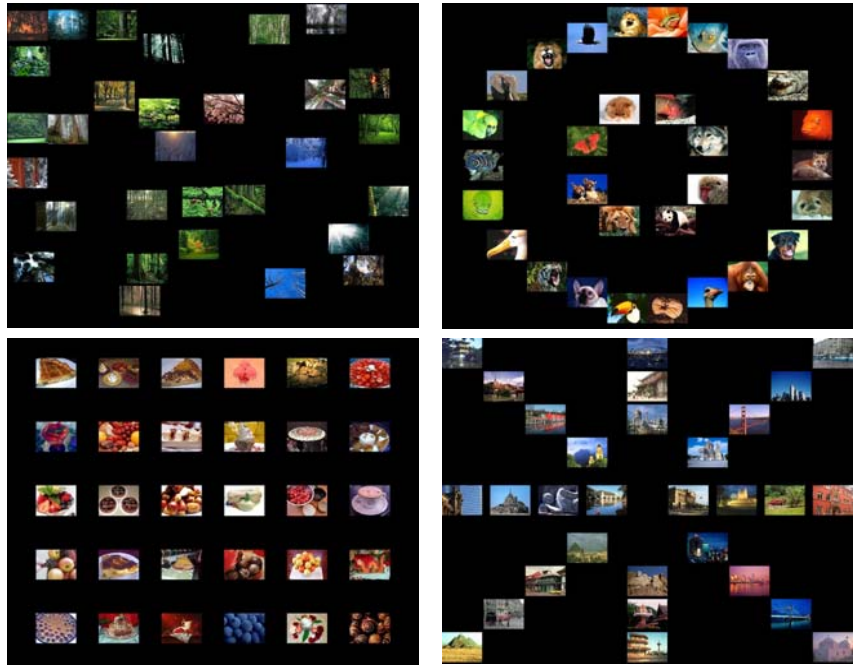

**Figure 1.** *Exemples de structures aléatoire, elliptique, matricielle et radiale.*

**Caractéristiques globales des scènes**

Nous avons retenu comme caractéristiques des scènes, d'une part, leur densité visuelle, caractérisée par le nombre de photographies de chaque collection, d'autre part, leur structure spatiale (aléatoire, elliptique, matricielle ou radiale). Toutes les scènes comprennent le même nombre de photographies, afin d'assurer la pertinence des comparaisons entre les quatre structures testées.

**Caractéristiques des collections d'images**

Pour étudier l'influence de la structure spatiale des affichages sur l'efficacité du repérage visuel, il est nécessaire de contrôler avec précision les paramètres visuels qui caractérisent chaque scène, donc les caractéristiques des collections de photographies utilisées. Chaque scène est composée de N photographies couleur, toutes au format 4/3, toutes de même taille, et disposées selon l'une des quatre structures retenues. De plus, pour que les photographies d'une même scène ne présentent pas entre elles des différences de saillance visuelle et de contenu sémiotique trop marquées, nous avons imposé qu'elles portent toutes sur le même thème. La saillance d'un item autre que la cible, en attirant le regard, pouvait ralentir la détection de celle-ci, tandis que la saillance de la cible pouvait faciliter et accélérer sa détection. Pour renforcer l'homogénéité visuelle interne des scènes par rapport aux paramètres forme, contrastes et niveau de détail de la saillance visuelle, nous avons distingué les photographies dites d'objets (objets complexes, personnages, ou groupes de personnages) des photographies de paysages ou de scènes d'intérieur.

Certains des travaux sur la recherche visuelle évoqués dans (Chelazzi, 1999), ont particulièrement retenu notre attention car ils montrent que le temps nécessaire à l'analyse perceptive de chaque item d'une scène croît lorsque la complexité des items augmente, ou lorsque la cible est similaire à des items non cibles (Treisman et Gormican, 1988 ; Duncan et Humphreys, 1989). Autrement dit, les items non cibles peu complexes et très différents de la cible sont rejetés plus rapidement que les items qui partagent une, voire plusieurs propriétés visuelles avec la cible. Ces résultats scientifiques nous ont conduits à distinguer trois niveaux de difficulté parmi les tâches de repérage, en fonction des caractéristiques visuelles des photographies de chaque collection :

- Niveau 1 dit « facile » : il se compose de collections de photographies hétérogènes visuellement, dans lesquelles la cible devrait être facile à détecter grâce aux mécanismes parallèles pré-attentifs qui permettent de « rejeter en bloc » un, voire plusieurs items non cibles.
- Niveau 2, de difficulté dite « moyenne » : il comprend des collections homogènes (visuellement) de photographies peu complexes (faible niveau de détail, contrastes importants) ; l'homogénéité visuelle devrait accroître les temps de repérage des cibles.
- Niveau 3 dit « difficile » : il est constitué de collections homogènes de photographies complexes (niveau de détail élevé) ; les temps de détection des cibles devraient être nettement plus élevés que pour les deux autres niveaux.

En ce qui concerne le niveau de détail des photographies, nos hypothèses sont compatibles avec le modèle de perception visuelle « coarse to fine » qui suppose que les fréquences spatiales basses (donc la structure globale d'une image) sont perçues plus rapidement que les hautes fréquences qui correspondent aux détails (Huges *et al.*, 1996).

A noter, comme l'illustre la figure 1, que nous avons sélectionné uniquement des photographies de bonne qualité visuelle, et veillé à ne regrouper dans une même collection que des photographies similaires en termes de contraste, luminosité et définition. Enfin, toutes les photographies sont sans contours ni bordures.

### *Caractéristiques des cibles*

Nous avons retenu, comme caractéristiques des cibles, leur position dans l'affichage et leur saillance visuelle dans la scène. La position des cibles dans la scène a été systématiquement variée pour occuper les 30 positions possibles par structure, afin de contrôler et neutraliser l'influence de ce facteur sur les performances des sujets, cette influence pouvant masquer les effets de deux variables expérimentales, l'assistance orale au repérage et la structure des affichages.

Nous avons également contrôlé, pour les raisons suivantes, la saillance visuelle relative de chaque cible par rapport aux items de la collection à laquelle elle appartient, appelée dans la suite « saillance interne ». Dans le cadre de notre première étude (Carbonell et Kieffer, 2002 et 2005), nous avons observé que la facilité de repérage d'une cible tient à trois facteurs principaux : sa saillance visuelle, sa similarité avec d'autres éléments de la scène, source de confusions, et son degré de familiarité pour les sujets (i.e., le caractère familier ou insolite du sujet représenté). Par conséquent, toutes les photographies d'une collection, y compris la cible, ont même forme (format 4/3) et même taille. Elles relèvent toutes d'un même thème, ce qui limite les différences de saillance visuelle entre elles. Pour caractériser les couples (cible, collection) en fonction de la difficulté du repérage de la cible, nous avons pris en compte, outre la saillance de la cible, sa similarité visuelle avec d'autres photographies de la collection dont elle fait partie. Plus le niveau de difficulté de la collection augmente (voir le paragraphe précédent), plus le nombre de confusions possibles entre la cible et des non cibles est élevé, en raison de l'homogénéité croissante de la collection et de la complexité grandissante des photographies qui la composent. le degré de familiarité des sujets avec les cibles a été aussi contrôlé : toutes les collections portent sur des thèmes de la vie courante :

animaux, avions, portraits ou groupes de personnages pour les photographies de type « objet complexe », bâtiments, montagnes, couchers de soleil ou scènes d'intérieur pour celles de type « paysage ».

Les collections ont été constituées manuellement et les cibles sélectionnées par deux experts humains. Nous n'avons pas utilisé, pour contrôler la saillance interne des items, le logiciel développé à partir du modèle d'attention visuelle proposé dans (Itti et Koch, 1999, 2000) en raison de ses limites. Si ce logiciel (Itti, 2000) est efficace pour l'analyse de scènes naturelles en deux dimensions, des développements logiciels importants auraient été nécessaires pour l'adapter au traitement de scènes telles que celles reproduites dans la figure 1. L'utiliser pour contrôler la saillance interne du contenu des photographies de chaque collection pour les propriétés forme et direction, aurait été très difficile en pratique et aurait fourni des résultats médiocres, comparés à ceux d'une évaluation humaine. En outre, la forme des items étant contrôlée (format 4/3) dans chaque scène, ainsi que leur intensité lumineuse, la seule propriété dont le logiciel aurait permis de contrôler la saillance interne aurait été les couleurs, qui sont faciles à évaluer par un sujet humain doué d'une vision normale des couleurs.

### 4.4 Caractéristiques du matériel sonore

La structure syntaxique des messages oraux est très simple ; elle se réduit à une courte expression spatiale précisant la zone de l'écran (parmi neuf zones pré-définies) dans laquelle la cible est située. Neuf messages ont été utilisés : « En haut à gauche. », « En haut. », « En haut à droite. », « A gauche. », « Au centre. », « A droite. », « En bas à gauche. », « En bas. » et « En bas à droite. ». Ces expressions correspondent à un découpage de l'écran en neuf zones rectangulaires de même surface. Les images chevauchant plusieurs zones sont affectées à la zone qui couvre la plus grande partie de leur surface. Les messages ont été lus par une locutrice expérimentée s'exprimant avec une prosodie « neutre », et enregistrés au format WAV. Ce découpage présente un double intérêt : d'une part, il se verbalise de façon très simple, on peut donc peut supposer qu'il est « naturel », d'autre part, il est compatible avec celui qu'utilisent les techniques actuelles de composition en photographie (Cadet *et al.*, 2002).

## 5 Protocole expérimental

### 5.1 Les variables de l'expérience

Pour mémoire, nous avons retenu comme caractéristiques des scènes, leur structure spatiale (4 structures) et leur densité en nombre d'éléments (N). Les collections de photographies qui les composent ont été classées en fonction de la difficulté du repérage de la cible (3 niveaux de difficulté) et du sujet des photographies (objet ou paysage). Les caractéristiques retenues pour les cibles sont leur saillance visuelle interne (contrôlée) et leur position dans la scène (occupation de toutes les positions possibles pour neutraliser l'influence éventuelle de ces paramètres sur les performances des sujets). Les neuf messages d'aide à la localisation de la cible dans la scène sont constitués d'indications spatiales absolues. Les variables libres étudiées sont les suivantes :
- le mode de présentation de la cible, visuel (PV) ou multimodal (PM), qui caractérise les deux situations ou conditions expérimentales considérées ;
- la structure de la scène : matricielle, radiale, elliptique, ou aléatoire ;
- le sujet des photographies : objet ou paysage ;
- le niveau de difficulté des collections de photographies : 1 (facile), 2 (moyen) ou 3 (difficile).

Les variables liées sont, comme dans notre première étude :
- le temps de sélection de la cible en millisecondes ;
- la précision de la sélection, à savoir, dans la cible ou en-dehors.

### 5.2 La validité interne

Nous avons assuré la validité interne du protocole expérimental en utilisant différentes techniques de contrôle : l'appariement et le contre-balancement des conditions expérimentales, l'affectation aléatoire des sujets, l'automatisation de la collecte des données et un questionnaire post-expérimental.

***Appariement des conditions expérimentales***

Afin d'obtenir des mesures comparables entre les deux conditions expérimentales, PV et PM, tout en neutralisant les effets possibles des différences interindividuelles, les sujets effectuent la tâche de repérage de cibles dans les deux conditions, de sorte qu'on observe dans chacune d'elles exactement le même niveau de variables nuisibles.

***Contre-balancement***

Nous avons fait varier l'ordre de présentation des conditions expérimentales. La moitié des sujets a été affectée à l'ordre PV-PM, l'autre moitié à l'ordre PM-PV. Le contre-balancement des conditions expérimentales permet d'éviter l'effet de séquence.

***Affectation aléatoire des conditions***

Nous avons ajouté à l'appariement et au contre-balancement, l'affectation aléatoire de l'ordre de présentation des scènes dans les deux conditions PV et PM et pour chaque sujet. Cette distribution aléatoire permet l'obtention de conditions équivalentes pour les différentes variables de l'expérimentation.

***Automatisation de la collecte des données***

Nous avons automatisé la collecte des données pour éviter les erreurs d'enregistrement. En effet, il existe toujours des comportements à la limite des définitions ou catégorisations utilisées dans le cadre d'une étude expérimentale (e.g., au cas particulier, les erreurs de sélection de la cible d'origine motrice). Il est à craindre que l'expérimentateur ait tendance à classer ces comportements dans le sens de son hypothèse de travail.

***Questionnaire post-expérimental***

Enfin, nous avons élaboré un questionnaire post-expérimental pour évaluer comment les participants ont perçu l'expérience. En effet, les activités cognitives peuvent interagir avec les procédures expérimentales et fausser les résultats. Les exigences implicites du protocole, comme la présentation de l'expérience, le rôle et l'objectif qu'on feint de lui donner (ici, évaluer une interface 2D interactive), le dispositif, ou encore les tâches à effectuer, définissent le point de vue du participant sur l'expérience, sur sa finalité. Dans le questionnaire post-expérimental, les participants peuvent livrer leur perception des objectifs de l'expérimentation, de ce qu'on attendait d'eux, du comportement qu'ils devaient adopter.

### 5.3 La validité échantillonnale

Assurer la validité échantillonnale de notre expérimentation a consisté à tenir compte à la fois du nombre de sujets participant à l'expérience, mais aussi du nombre de scènes qui leur étaient présentées. En effet, la difficulté à généraliser les résultats de notre première étude (Carbonell et Kieffer, 2002 et 2005) ne résidait pas tant dans le nombre de participants (18) à l'expérience ni dans le nombre de scènes présentées (36) – des travaux au protocole expérimental similaire (18 sujets, 3 conditions expérimentales, 7 tâches par condition) ont été validés par leur publication dans les actes de la conférence internationale CHI'92 (Ahlberg *et al.*, 1992) – mais plutôt dans le fait de ne pouvoir comparer les performances de chaque sujet sur la même image dans les deux conditions expérimentales, en raison du nombre limité d'images par condition (12) et du nombre réduit de sujets par image et par condition. Nous avons donc fixé à 24 le nombre de participants et imposé qu'ils appartiennent à une même catégorie homogène d'utilisateurs potentiels habitués à l'utilisation de la souris, de façon à contrôler les phénomènes d'apprentissage du mode de sélection des cibles au cours de l'expérimentation. Nous avons choisi 22 étudiants, élèves ingénieurs et doctorants en informatique. Les deux autres sujets sont des informaticiens (BAC+4), employés d'entreprises nancéiennes. Les sujets étaient donc tous des utilisateurs experts de la souris compte tenu de leur activité principale. Nous avons choisi une tranche d'âge étroite, entre 24 et 29 ans, de façon à limiter la variabilité inter-individuelle, surtout au niveau des comportements sensori-moteurs. En effet, des études en psychologie (Dollinger et Hoyer, 1996) ont montré que les performances en termes de précision et de temps de réponse, lors de tâches d'inspection visuelle, varient avec l'âge. Enfin, nous avons vérifié (à l'aide de tests visuels classiques) que les sujets avaient tous une vision normale.

Nous avons fixé le nombre de scènes par condition expérimentale à 120, soit 30 scènes par structure, 40 par niveau de difficulté et 60 par type de sujet des photographies (objet ou paysage). Autrement dit, pour chaque structure, on dispose, dans chaque condition expérimentale, de 30 scènes comprenant 10 scènes par niveau de difficulté et 15 par type de sujet. Enfin, nous avons fixé le nombre N de photographies par scène à 30, et donc à 30 également le nombre de positions possibles pour la cible par condition et par structure. Un nombre plus élevé de photographies par scène risquait d'entraîner un échec systématique du repérage de la cible dans les scènes de difficulté 2 et 3.

La réalisation d'un plan expérimental permettant de comparer en toute rigueur les performances des sujets vis-à-vis des quatre structures d'affichage des scènes et des deux modes de présentation de la cible aurait imposé de définir 30 couples (scène, cible) et de les présenter huit fois aux sujets, protocole adopté par exemple dans (Van Diepen *et al.*, 1999). Mais cette solution aurait rendu l'activité de repérage artificielle et fastidieuse. Elle risquait donc de créer chez les sujets des effets de lassitude et de démotivation en cours d'expérimentation. En outre, elle ne permet pas, même en augmentant considérablement le nombre de scènes, de garantir l'absence de phénomènes d'apprentissage de la tâche. Nous avons donc choisi le compromis suivant, qui tient compte des objectifs principaux de cette étude en privilégiant les comparaisons entre les deux modes de présentation PV et PM : les mêmes 120 scènes associées aux mêmes cibles sont présentées dans les deux conditions, mais dans des ordres différents définis aléatoirement. Pour mémoire, l'ordre de passation est contrebalancé, un groupe de 12 sujets traitant les conditions expérimentales dans l'ordre PV-PM, l'autre dans l'ordre PM-PV. Par ailleurs, pour que l'ordre de présentation des scènes n'influence pas les résultats en formant des séries d'images plus faciles/difficiles que d'autres, nous avons fait le choix de définir aléatoirement l'ordre de présentation des scènes pour chaque sujet. Enfin, les 120 scènes illustrant les quatre structures sont toutes entièrement différentes (i.e., les photographies qui les composent sont toutes différentes).

### 5.4 Développements logiciels réalisés

La mise en œuvre du protocole expérimental adopté imposait, par souci de rigueur et d'efficacité, d'automatiser :

- La création des 120 scènes présentées aux sujets, en particulier la disposition des photographies dans les différentes structures. Les 3600 photographies utilisées ont été sélectionnées et triées manuellement pour constituer les 120 collections nécessaires ; elles provenaient d'une base de données que nous avons créée et qui contenait 6000 photographies extraites de sites Web populaires.
- Le déroulement des passations (24 sujets), notamment l'ordre d'affichage des scènes pour chaque sujet, et le recueil des performances de chacun d'eux (temps de réponse et positions des clics de sélection des cibles).

Les développements logiciels ont été effectués en Java.

### 5.5 Déroulement d'une passation

Le déroulement d'une passation est le même pour tous les sujets, à l'exception de l'ordre de passation des conditions expérimentales (PV et PM) qui varie d'un groupe à l'autre, et de l'ordre de présentation des scènes, différent pour chaque sujet. La durée globale d'une passation est d'environ 1h30 et comprend, dans l'ordre, les étapes suivantes :

- Lecture de la présentation générale de l'expérience et remplissage d'un questionnaire d'identification (environ 5 minutes).
- Tests de vision réalisés par l'expérimentateur (environ 15 minutes).
- Présentation des consignes par l'expérimentateur, puis court entraînement initial en sa présence (environ 5 minutes).
- Réalisation des 120 tâches de repérage dans une condition, puis remplissage d'un questionnaire d'évaluation portant sur cette condition (environ 25 minutes).
- Réalisation des 120 tâches de repérage dans l'autre condition, puis remplissage d'un questionnaire d'évaluation portant sur cette condition et sur la comparaison des deux conditions (environ 25 minutes).
- Entretien final de débriefing (entre 10 et 20 minutes).

*Tests de vision*

Chaque sujet a effectué des tests portant sur la vision binoculaire et monoculaire, de près et de loin (utilisation des tests de vision BIOPTOR mis au point par Stereo Optical Co., Inc.). Tous les participants à l'expérimentation présentent une vue normale en termes d'acuité visuelle, de fusion latérale et de fusion centrale de près, de discrimination des couleurs. Quant à leur vision de loin elle est dénuée d'anomalie.

*Entraînement des sujets*

La familiarisation avec la tâche, qui suit la lecture des consignes, porte sur dix scènes, cinq par situation, dans l'ordre PV-PM ou PM-PV, selon que le sujet passe l'expérimentation dans l'ordre PV-PM ou PM-PV. Le rôle de l'expérimentateur est limité au commentaire de la consigne et à la présentation des tâches d'entraînement. Il assiste à leur réalisation, puis répond aux éventuelles questions du sujet. Il quitte la salle pendant la passation, mais est disponible par téléphone. Le sujet dispose en permanence d'une version écrite de la consigne et du numéro où il peut contacter l'expérimentateur. Les entretiens de débriefing ont été assurés par un autre chercheur de l'équipe.

*Questionnaires*

Le premier questionnaire concerne le profil des participants : niveau de culture général (diplôme en cours de préparation ou profession) et compétences en informatique (Internet, applications grand public, logiciels graphiques, jeux, programmation, etc.).

Les deux autres questionnaires concernent l'évaluation subjective des modes de présentation de la cible et des structures d'affichage. Les questions [1], identiques pour les deux conditions, portent sur l'attrait et la difficulté des tâches de repérage, les performances réalisées (rapidité, précision), la fatigue éventuellement ressentie. Le dernier questionnaire invite en outre les sujets, d'une part, à évaluer l'efficacité relative des quatre structures d'affichage et préciser celle qu'ils préfèrent et, d'autre part, à comparer les deux modes de présentation des cibles.

## 6   Résultats

Les analyses statistiques ont été réalisées en collaboration avec François-Xavier Jollois, Maître de Conférences à l'Université René Descartes (Paris V). Elles consistent essentiellement en des tests t portant sur la comparaison entre les deux modes de présentation des cibles (PV et PM), les quatre structures des affichages (aléatoire, elliptique, matricielle et radiale), les trois niveaux de difficulté des tâches de repérage, et le sujet des photographies (objets versus paysages). Les données ont été appariées par sujet (paired t-tests).

### 6.1   Présentation multimodale *versus* présentation visuelle

Afin de comparer les performances des sujets dans les deux conditions expérimentales, nous avons regroupé les données recueillies par condition (cf. les tableaux 1 et 2).

| | Temps de sélection des cibles | | |
|---|---|---|---|
| Condition | Moyenne (ms) | Ecart type (ms) | Nb. observations |
| PV | 5674 | 5985 | 2880 |
| PM | 1747 | 1552 | 2880 |

**Tableau 1.** *Résultats globaux : temps moyens de sélection des cibles par condition.*

| | Précision de la sélection des cibles | | |
|---|---|---|---|
| Condition | Nombre d'erreurs | Taux d'erreurs (%) | Nb. observations |
| PV | 150 | 5.2 | 2880 |
| PM | 79 | 2.7 | 2880 |

**Tableau 2.** *Résultats globaux : précision de la sélection des cibles par condition. Les pourcentages ont été calculés sur l'ensemble des tâches (120) par condition.*

### *Temps de sélection des cibles*

Le temps moyen de sélection des cibles observé dans la condition visuelle PV (5674 ms), associé à un écart type élevé (5985 ms), est plus de 3 fois supérieur à celui observé dans la condition multimodale (1747 ms), associé à un écart type faible (1552 ms). En d'autres termes, les sujets sont 3 fois plus rapides et beaucoup plus réguliers dans la condition multimodale qu'ils ne le sont dans la condition visuelle. Ce résultat est statistiquement hautement significatif (t=-34.07 ; p<.0001).

---

[1] Utilisation de différenciateurs sémantiques (Osgood). La technique des différenciateurs sémantiques associe à un objet/concept donné (e.g., « tâche »), un ensemble de paires d'adjectifs à significations clairement opposées (e.g., « facile », « difficile »). Chaque paire est présentée aux extrémités d'une échelle comprenant le plus souvent 5 ou 7 cases ; les participants cochent la case exprimant le mieux leur jugement.

### *Précision de la sélection des cibles*

Globalement, le taux d'erreurs est peu élevé (moins de 6% dans les deux conditions expérimentales). A noter cependant que, dans la condition visuelle (PV), il est près de 2 fois supérieur à celui observé dans la condition multimodale (PM).

### *Analyse des questionnaires*

Les 24 sujets se jugent plus rapides dans la condition multimodale qu'ils ne pensent l'être dans la condition visuelle. Ce qui est conforme à la réalité puisque, quel que soit l'ordre de passation, tous les sujets sont plus rapides dans la condition multimodale.

En ce qui concerne la facilité du repérage, 75% des sujets (18 sujets) ont jugé les tâches de repérage plus faciles dans la condition multimodale ; un seul sujet, qui a effectué les tâches dans l'ordre PM-PV, a estimé le repérage plus facile dans la condition visuelle. Quatre sujets n'ont pas ressenti de différence entre les deux conditions, et le dernier s'est déclaré sans avis sur la question. De nombreux sujets ont précisé que la condition multimodale réduisait les hésitations et les « clics au hasard » estimés plus fréquents dans la condition visuelle. La difficulté supérieure du repérage visuel en l'absence d'indications sur la position de la cible pourrait donc expliquer les différences de performances des sujets entre les deux conditions.

Une large majorité des sujets a préféré effectuer les tâches de repérage dans la condition multimodale (16 sujets). Les autres ont déclaré avoir été plus intéressés par la condition visuelle car la tâche leur a paru plus « amusante » et/ou « stimulante » dans cette condition (cf. les entretiens de débriefing). Mais tous ont considéré les messages comme une assistance efficace au repérage de cibles ; ils n'ont pas été déroutés ni gênés par ces indications orales.

Enfin, 16 sujets considèrent les tâches de repérage visuel comme familières. Aucun des huit autres sujets n'a été gêné par leur caractère inhabituel.

### *Interprétation et conclusions : effets du mode de présentation de la cible*

D'après l'analyse globale des performances des sujets et de leurs jugements subjectifs sur les deux conditions expérimentales, la présentation multimodale de la cible facilite et améliore de façon très importante le repérage visuel par rapport à sa présentation visuelle, en termes du temps de sélection (résultat statistiquement significatif) et de la précision de la sélection de la cible. Ces résultats valident l'hypothèse H1.

D'après les commentaires spontanés recueillis lors des débriefings, il semble que, dans la condition multimodale, les sujets dirigent leur regard vers la zone indiquée oralement, dès l'apparition de la scène. En effet, certains sujets ont précisé qu'ils avaient adopté deux stratégies différentes en fonction du mode de présentation de la cible. Selon eux, dans la condition multimodale, ils suivent simplement l'indication contenue dans le message. Dans la condition visuelle, en revanche, ils tentent soit de se construire et de mémoriser une description verbale de la cible, soit de mémoriser un détail qu'ils espèrent discriminant. Au moment où la scène apparaît, ils la parcourent systématiquement, souvent en s'appuyant sur sa structure visuelle. Il semble donc que les sujets aient effectivement utilisé les messages oraux pour réduire la zone de recherche de la cible et qu'ils aient tiré parti de l'information spécifique véhiculée par chacune des modalités. La réduction importante des temps de sélection et des taux d'erreurs observés dans la condition multimodale par rapport à la condition visuelle peut donc vraisemblablement s'expliquer par l'influence des messages sur les stratégies de parcours visuels des sujets. A noter que les commentaires formulés par certains sujets sur la nécessité de se créer une représentation mentale consciente (donc verbalisable) de la tâche à réaliser sont en accord avec ceux décrits dans (Pelz *et al.*, 2001) pour des tâches simples de manipulation de blocs où la perception visuelle guide le geste.

L'analyse des questionnaires et des débriefings montre également que les sujets ont jugé les messages utiles et qu'ils les considèrent comme une « assistance efficace à la tâche ». Le fait que les informations spatiales soient exprimés oralement ne les a pas « gênés ». En outre, ces messages ne semblent pas avoir augmenté leur charge cognitive, bien au contraire, puisque la plupart d'entre eux ont indiqué avoir ressenti moins de fatigue lors de la condition multimodale que lors de la condition visuelle qu'ils jugent plus difficile et moins rapide. La fusion des deux types d'informations, présentation visuelle de la cible et indications orales sur sa position dans la scène, ne semble pas avoir surchargé leur mémoire de travail. Des résultats similaires issus de la psychologie cognitive sont présentés dans (Kalyuga *et al.*, 1999). Les participants à cette étude devaient effectuer des exercices interactifs d'apprentissage sur les techniques de soudure dans l'une des trois conditions expérimentales suivantes, qui ne différaient entre elles que par les modalités d'expression des connaissances à acquérir : dans la première, un texte présenté sous forme écrite et orale accompagnait un diagramme, dans la seconde et la troisième, ce texte était présenté uniquement sous l'une des deux formes, écrite ou orale.

### 6.2 Effets de l'ordre de passation

#### *Deux groupes de sujets, deux stratégies de recherche différentes*

Quelle que soit la condition expérimentale considérée, les sujets du groupe 2 (ordre de passation PM-PV) se sont montrés plus rapides que les sujets du groupe 1 (ordre PV-PM). Dans la condition visuelle, la différence moyenne observée entre les deux groupes est de 2522 ms. Ce résultat est statistiquement significatif ($t=11.56$ ; $p<.0001$). Dans la condition multimodale, la différence moyenne observée entre les deux groupes est de 213 ms. Bien que cet écart entre les deux groupes soit faible, il reste statistiquement significatif ($t=3.7$ ; $p=.0002$).

En outre, toutes conditions confondues, les sujets du groupe 1 effectuent les tâches de repérage avec une précision de 99% (93 erreurs ; 2880 observations) tandis que les sujets du groupe 2 les effectuent avec une précision de 95.5% (136 erreurs ; 2880 observations). Cette différence est statistiquement significative ($t=5.16$ ; $p<.0001$). Ces résultats suggèrent que le groupe 1 adopte une stratégie privilégiant la précision de la sélection des cibles, tandis que le groupe 2 privilégie la rapidité de sélection des cibles.

#### *Repérage visuel des cibles et apprentissage*

Nous avons analysé l'évolution des performances des sujets par groupe (1 et 2) et par condition expérimentale (PV et PM). Les analyses qui suivent reposent sur l'interprétation des graphiques A1, A2, A3 et A4 en annexe. Ils illustrent l'évolution des performances des sujets sur des paquets de 30 images (ou scènes) : images 1 à 30 ; 31 à 60 ; 61 à 90 ; 91 à 120. Les observations ont été regroupées par groupe de sujets et par condition expérimentale.

D'après le graphique A1, l'évolution des temps de sélection des cibles pendant la condition visuelle représente, pour le groupe 1, l'évolution classique de l'apprentissage d'une nouvelle tâche, compte tenu du temps limité de la passation. En effet, on observe une diminution des temps de sélection sur les 60 premières images, puis une augmentation, due probablement à la fatigue, au relâchement des efforts (confiance en soi et assurance excessives) et/ou à une baisse de la motivation. Les temps de sélection des cibles du groupe 2 pendant la condition visuelle évoluent différemment : il semble que les sujets aient sous-estimé la difficulté de la tâche sur les 60 premières images (sous l'influence de la condition multimodale réalisée précédemment), puis qu'ils se soient ressaisis, mais que la fatigue éprouvée (il s'agit de la seconde condition expérimentale) ait ensuite interféré avec leurs efforts et limité l'amélioration de leurs performances sur les 60 dernières images. Cette « contre-performance » du groupe 2 peut également provenir, en partie, du délai nécessaire pour que les sujets perçoivent l'identité des scènes entre les deux conditions.

D'après le graphique A2, l'évolution des temps de sélection des cibles pendant la condition multimodale est semblable, pour le groupe 1, à celle obtenue pour le groupe 2 dans la condition visuelle. La difficulté à faire évoluer la stratégie élaborée au cours de la première condition pour l'adapter à une activité de recherche visuelle apparemment semblable, mais en fait très différente (voir 6.1), peut rendre compte de leurs performances sur les 60 premières images, tandis que la fatigue peut expliquer l'amélioration limitée de celles-ci sur les 30 dernières. L'évolution des temps de sélection du groupe 2 dans la condition multimodale est caractéristique de l'apprentissage d'une nouvelle tâche. Toutefois, les effets sont moins marqués que pour le groupe 1 dans la condition visuelle ; en particulier, il n'y a pas d'augmentation des temps moyens de sélection des cibles sur les 60 dernières images, vraisemblablement parce que la tâche est moins exigeante, donc moins fatigante que dans la condition visuelle.

D'après le graphique A3, concernant la précision de la sélection des cibles dans la condition visuelle, il n'y a pas d'évolution notable des erreurs pour le groupe 1, vraisemblablement parce que les sujets privilégient la précision par rapport à la rapidité. La courbe du groupe 2 dans la condition visuelle représente l'évolution classique de l'apprentissage d'une nouvelle tâche relativement facile (les cibles et les images ont déjà été examinées une première fois) : on n'observe ni fatigue, ni relâchement des efforts, le nombre des erreurs décroît régulièrement.

D'après le graphique A4, concernant la précision de la sélection des cibles dans la condition multimodale, la courbe du groupe 1 met en évidence une augmentation sensible du nombre d'erreurs sur les 30 dernières images (par rapport aux images 60-90), due à la fatigue vraisemblablement. Les sujets du groupe 1 sont plus fatigués car « leur » condition visuelle s'est avérée plus difficile que celle du groupe 2 puisque, dans cette condition, ils voyaient les cibles et les scènes pour la première fois. La courbe d'évolution du nombre d'erreurs du groupe 2 dans la condition multimodale ressemble à celle d'un apprentissage d'une nouvelle tâche. Cependant, comment expliquer l'augmentation importante du nombre d'erreurs sur les images 30 à 60 ? Cette régression passagère des performances peut être due à un excès d'assurance : croyant avoir maîtrisé cette tâche relativement facile, les sujets du groupe 2 se précipitent et commettent beaucoup plus d'erreurs que le groupe 1.

Ces résultats sont intéressants car ils renseignent sur la fatigue associée à chacune des deux formes de recherche visuelle, avec ou sans l'assistance d'informations sur la position de la cible ; voir les différences d'évolution des temps moyens de sélection sur les images 60 à 120 entre les deux conditions et les deux groupes de sujets. En particulier, la comparaison des performances des sujets en fonction de l'ordre d'exécution des deux conditions montre que, en l'absence d'indications sur la position de la cible dans la scène, le repérage visuel est une activité plus exigeante et plus fatigante qu'en la présence de telles informations.

### 6.3 Influence de la structure des affichages sur les performances des sujets

L'analyse de l'influence de la structure des affichages sur les performances des sujets se découpe en deux volets : une analyse statistique qui porte sur la variable temps de sélection des cibles et une étude qualitative qui porte sur l'analyse des erreurs. L'objectif que nous nous fixons ici est de tester l'hypothèse de travail H2, c'est-à-dire de déterminer s'il existe des différences significatives en termes de temps et de précision de la sélection des cibles entre les quatre structures testées, avec ou sans assistance orale au repérage de la cible.

Afin de mettre en évidence l'éventuelle influence de l'organisation spatiale des scènes sur les performances des sujets, nous avons comparé la rapidité et la précision de la sélection des cibles entre les quatre structures testées, dans chacune des deux conditions expérimentales. Les données recueillies lors de l'expérience ont été regroupées par condition et par structure (aléatoire, elliptique, matricielle et radiale) ; on obtient ainsi 8 ensembles de 720 observations. Une donnée est un couple (temps de sélection ; précision de la sélection).

**Temps de sélection des cibles**

Les temps moyens de sélection et les écarts types, en millisecondes, par structure et par condition expérimentale, sont présentés dans le tableau 3.

| Structure | Condition | Moyenne (ms) | Ecart type (ms) |
|---|---|---|---|
| Aléatoire | PV | 5626 | 5819 |
| | PM | 1737 | 1437 |
| Elliptique | PV | 6250 | 6585 |
| | PM | 1851 | 1633 |
| Matricielle | PV | 5738 | 5879 |
| | PM | 1763 | 1819 |
| Radiale | PV | 5081 | 5565 |
| | PM | 1640 | 1256 |

**Tableau 3.** *Temps moyens de sélection des cibles par structure et par condition.*
*(720 observations par structure et par condition)*

Dans la condition PV, le classement des structures par ordre croissant du temps moyen de sélection des cibles est le suivant :
- structure radiale, avec un temps moyen de 5081 ms ;
- structures aléatoire et matricielle, avec des temps moyens voisins (5626 ms et 5738 ms, respectivement) ;
- structure elliptique, avec un temps moyen de 6250 ms.

Les différences observées dans la condition PV entre les temps moyens de sélection des cibles par structure sont statistiquement significatives pour :
- les structures elliptique et radiale (t=3.64 ; p=.0003),
- les structures matricielle et radiale (t=2.18 ; p=0.0296) et
- les structures elliptique et matricielle (t=1.25 ; p=0.0024).

On observe une tendance entre, d'une part, la structure aléatoire et, d'autre part, les structures elliptique et radiale : (t=-1.91 ; p=0.0568) et (t=1.82 ; p=0.0696), respectivement. La différence entre les structures aléatoire et matricielle, en revanche, n'est pas significative (t=-0.30 ; p=0.71).

Dans la condition PM, les structures se classent comme suit, par ordre croissant du temps moyen de sélection des cibles :
- structure radiale, avec un temps moyen de 1640 ms ;
- structures aléatoire et matricielle, avec des temps moyens respectifs de 1737 et 1763 ms ;
- structure elliptique, avec un temps moyen de 1851 ms.

Les différences observées dans la condition PM entre les temps moyens de sélection des cibles par structure sont statistiquement significatives pour les structures elliptique et radiale uniquement (t=2.75 ; p=0.006). On observe une tendance entre les structures matricielle et radiale (t=1.49 ; p=0.1352), et entre les structures aléatoire et radiale (t=1.37 ; p=0.17). On note également une tendance entre les structures aléatoire et elliptique (t=-1.40 ; p=0.16). Les différences entre, d'une part, les structures aléatoire et matricielle, d'autre part, les structures elliptique et matricielle, ne sont pas significatives.

### Précision de la sélection des cibles

Dans le tableau 4 figurent, les nombres d'erreurs effectives par structure et par condition expérimentale, ainsi que leur répartition sous forme de pourcentages calculés par rapport au nombre total d'erreurs effectives par condition, soit 96 et 49 pour les conditions PV et PM respectivement. Les erreurs dites effectives désignent les sélections erronées (i.e., les sélections de non cibles) à l'exclusion des sélections imprécises (au voisinage de la cible) et des échecs de la recherche visuelle (clics sur le fond de l'affichage, à distance de toute photographie).

| Précision de la sélection des cibles | | | |
|---|---|---|---|
| Structure | Condition | Nb. erreurs | Répartition (%) |
| Aléatoire | PV | 20 | 21 % |
| | PM | 13 | 27 % |
| Elliptique | PV | 25 | 26 % |
| | PM | 18 | 37 % |
| Matricielle | PV | 27 | 28 % |
| | PM | 12 | 24 % |
| Radiale | PV | 24 | 25 % |
| | PM | 6 | 12 % |

**Tableau 4.** *Nombre et taux des erreurs effectives par structure et par condition. Les pourcentages ont été calculés sur les nombres d'erreurs effectives par condition, soit 96 pour PV et 49 pour PM. (720 observations par structure et par condition)*

Dans la condition visuelle, on observe une répartition relativement équilibrée des erreurs entre les quatre structures. Les pourcentages d'erreurs sont compris entre 21 et 28%. Toutefois, la structure aléatoire se distingue des autres puisqu'elle entraîne un nombre d'erreurs inférieur à celui observé pour les autres structures.

Le résultat le plus intéressant concerne la condition multimodale. En effet, entraînant trois fois moins d'erreurs que la structure elliptique et deux fois moins d'erreurs que les structures matricielle ou aléatoire, la structure radiale est dans cette condition celle qui permet la détection de cibles la plus précise. Les indications spatiales absolues améliorent donc davantage la précision de la recherche visuelle lorsque la scène présente une structure radiale plutôt que l'une des trois autres structures, en particulier la structure elliptique. Le découpage induit par les indications spatiales verbales (i.e., 9 rectangles) peut expliquer ce résultat.

### Conclusions : effets de la structure des affichages

L'analyse des performances des sujets par condition et par structure met en évidence, dans la condition PM, des différences entre l'efficacité des structures étudiées, à la fois en terme de rapidité et de précision de la sélection des cibles. D'une part, concernant les temps de sélection moyens, on observe une différence moyenne de 211 ms entre la structure radiale et la structure elliptique, et ce résultat est statistiquement significatif. Ainsi, bien que les messages multimodaux tendent à niveler les différences entre les temps moyens de sélection par structure, l'organisation radiale est celle pour laquelle les sujets sont les plus rapides, et la structure elliptique celle pour laquelle ils sont les plus lents. D'autre part, concernant la précision de la sélection des cibles, on observe des différences importantes entre les structures : le nombre d'erreurs est multiplié par 3 entre la structure radiale (6 erreurs) et la structure elliptique (18 erreurs), et par 2 entre la structure radiale et les structures matricielle et aléatoire (12 et 13 erreurs, respectivement). Dans la condition PM, la structure radiale apparaît donc comme l'organisation spatiale la plus efficace. C'est pour cette structure que l'assistance à la recherche visuelle fournie par les messages oraux est la plus efficace, et pour la structure elliptique qu'elle est la plus limitée.

Les différences entre les effets des quatre structures sur l'efficacité du repérage visuel sont plus marquées dans la condition PV, pour les temps de sélection. En effet, on observe jusqu'à plus d'une seconde d'écart en moyenne (1169 ms) entre la structure elliptique et la structure radiale. La plus petite différence observée, entre la structure matricielle et la structure aléatoire (112 ms), est la seule qui ne soit pas statistiquement significative, contrairement aux autres différences qui expriment, dans le pire des cas, une tendance. La structure radiale est donc l'organisation spatiale pour laquelle les temps de sélection sont les plus courts dans les deux conditions. Cependant, dans la condition PV, aucune structure ne se distingue des autres quant à la précision du repérage : les erreurs se répartissent équitablement entre les quatre structures.

Par conséquent, l'ensemble des résultats présentés dans cette section contribuent à valider l'hypothèse de travail H2, dans la mesure où ils montrent que, dans les deux conditions expérimentales, la structure de l'affichage influence la rapidité d'exécution des tâches de repérage visuel proposées aux sujets, la structure radiale étant celle pour laquelle l'exécution des tâches s'est avérée la plus rapide. Cependant, ces résultats ne permettent de valider H2 que partiellement, car aucune conclusion ne peut être dégagée de l'analyse des erreurs, en raison de leur nombre global très faible et de l'absence d'influence notable de la structure spatiale sur la précision des sélections dans la condition PV. Pour étudier l'influence éventuelle de la structure spatiale des affichages sur la précision des sélection, il serait nécessaire d'augmenter sensiblement la difficulté des tâches de repérage visuel, de façon à augmenter la fréquence des erreurs.

Comme c'est souvent le cas, les performances des sujets sont en contradiction avec les préférences que la majorité d'entre eux ont exprimées dans les questionnaires et lors des débriefings. De même, on observe des divergences entre leurs jugements subjectifs écrits et oraux. En effet, une large majorité des sujets ont exprimé dans les questionnaires une nette préférence (indépendamment de la condition expérimentale) pour la structure elliptique, la moins efficace des quatre, et classé les structures par ordre décroissant de préférence comme suit : elliptique, radiale (la plus efficace des quatre structures), matricielle et aléatoire. Cependant, au cours des débriefings, les sujets ont nuancé leur jugement en fonction des situations d'interaction. 21 sur les 24 ont exprimé leur préférence, dans la condition multimodale, pour la structure radiale. Certains d'entre eux ont en outre formulé des jugements très critiques sur la structure elliptique qui leur semblait, contrairement à la structure radiale, mal adaptée au découpage spatial induit par les messages oraux et peu efficace, car « il y avait plus d'images à regarder » pour repérer la cible.

L'analyse détaillée des débriefings permet en outre d'expliquer de manière plausible pourquoi la structure matricielle, bien que la plus familière aux sujets (ne serait-ce qu'en raison de son utilisation pour l'affichage des icônes par de nombreux gestionnaires de fenêtres), a entraîné des parcours relativement lents et un nombre d'erreurs relativement élevé. La plupart des sujets ont déclaré au cours des débriefings avoir adopté, pour explorer les scènes matricielles, une stratégie de parcours linéaire (suivant les lignes et/ou les colonnes). D'autres ont précisé qu'ils n'avaient pas su par où commencer leur exploration des affichages matriciels. Ces remarques sont en accord avec les observations de Perkins (Perkins, 1995) concernant la navigation visuelle dans les pages des services en ligne.

### 6.4 Influence de la difficulté des scènes sur les performances des sujets

Afin de mettre en évidence l'influence éventuelle de la difficulté des scènes sur les performances des sujets, nous avons comparé la rapidité et la précision de la sélection des cibles en fonction des trois niveaux de difficulté considérés dans cette étude (voir 4.3). Les données sur les performances des sujets ont été triées par condition et par niveau de difficulté ; ce qui donne 6 ensembles de 960 observations chacun. Les résultats sont présentés dans les tableaux 5 et 6.

| Temps de sélection des cibles | | | |
|---|---|---|---|
| Difficulté | Condition | Moyenne (ms) | Ecart type (ms) |
| Facile | PV | 4919 | 5011 |
| | PM | 1611 | 1387 |
| Moyenne | PV | 5439 | 5879 |
| | PM | 1620 | 1272 |
| Difficile | PV | 6663 | 6801 |
| | PM | 2012 | 1893 |

**Tableau 5.** *Temps moyens de sélection des cibles par niveau de difficulté et condition. (960 observations par niveau de difficulté et par condition)*

| Précision de la sélection des cibles | | | |
|---|---|---|---|
| Difficulté | Condition | Nb. erreurs | Répartition % |
| Facile | PV | 17 | 18 |
| | PM | 10 | 20 |
| Moyenne | PV | 33 | 34 |
| | PM | 15 | 31 |
| Difficile | PV | 46 | 48 |
| | PM | 24 | 49 |

**Tableau 6.** *Nombre et taux des erreurs effectives par niveau de difficulté et condition. Les pourcentages ont été calculés sur les nombres d'erreurs effectives par condition, 96 pour PV et 49 pour PM. (960 observations par niveau de difficulté et par condition)*

**Temps de sélection des cibles**

On observe dans chacune des deux conditions des différences de rapidité entre les trois niveaux de difficulté. Pour la condition visuelle, ces différences importantes, surtout entre le niveau 3 et les deux autres niveaux, sont toutes statistiquement significatives (hautement significatives pour les deux dernières) : 520 ms entre les niveaux 1 et 2 (t=-2.07 ; p=0.0369), 1744 ms entre les niveaux 1 et 3 (t=-6.40 ; p<.0001), 1224 ms entre les niveaux 2 et 3 (t=-4.22 ; p<.0001). Pour la condition multimodale, la différence de 9 ms entre les niveaux 1 et 2 n'est pas significative. En revanche, les différences plus importantes, 401 ms et 392 ms, observées entre les niveaux 1 et 3, respectivement 2 et 3, sont hautement significatives (t=-5.29 ; p<.0001 et t=-5.33 ; p<.0001).

### *Précision de la sélection des cibles*

Le nombre des erreurs commises par les sujets, en fonction du niveau de difficulté des scènes et de la condition expérimentale, est présenté dans le tableau 6. On observe dans chacune des deux conditions des différences entre les niveaux de difficulté. Dans la condition visuelle, le nombre d'erreurs est multiplié par 2 entre les niveaux 1 et 2. Il est presque multiplié par 3 entre les niveaux 1 et 3. Les différences sont moindres dans la condition multimodale. A noter que dans chaque condition, le niveau de difficulté 3 est responsable de près de 50% des erreurs : 48% dans la condition visuelle, 49% dans la condition multimodale.

A noter également que, pour chaque niveau de difficulté, les erreurs sont moins nombreuses dans la condition PM que dans la condition PV. Par exemple, pour le niveau 2, le nombre d'erreurs est divisé par plus de 2 dans la condition multimodale : 33 erreurs dans PV versus 15 erreurs dans PM. On obtient des résultats voisins pour les deux autres niveaux de difficulté.

### *Conclusions : effets de la difficulté de la tâche*

Les résultats quantitatifs obtenus, temps de sélection et précision de la sélection, confirment les hypothèses sur lesquelles nous avons fondé la caractérisation des collections d'images et, par conséquent, les niveaux de difficulté des tâches de repérage visuel. Les sujets ont effectivement repéré plus rapidement et avec une meilleure précision les cibles, lorsque les collections de photographies étaient hétérogènes plutôt qu'homogènes et que les photographies composant la collection, y compris la cible, présentaient une faible complexité de détail.

En outre, les résultats que nous venons de présenter montrent que, plus le repérage de la cible est difficile, plus la présentation multimodale de celle-ci s'avère efficace. Par conséquent, ces résultats contribuent à démontrer l'efficacité d'une assistance à la recherche visuelle dans des affichages complexes qui se manifeste, dans des situations d'interaction exigeantes visuellement pour l'utilisateur, par des messages oraux constitués d'informations spatiales absolues sur la position approximative de la cible dans l'affichage. La comparaison, entre les deux conditions, des temps de sélection de la cible par niveau de difficulté conforte cette conclusion. Les différences observées entre les conditions PV et PM sont significativement plus importantes pour les scènes de niveau 2 que pour les scènes de niveau 1 (t=-2.07 ; p=0.0382), pour les scènes de niveau 3 que pour les scènes de niveau 1 (résultat hautement significatif, t=-4.92 ; p<.0001), et pour les scènes de niveau 3 que pour les scènes de niveau 2 (t=-2.90 ; p=0.0037).

Par conséquent, lors de la conception d'une application interactive de recherche visuelle, la décision de mettre en œuvre ou non des messages oraux d'assistance à la localisation de l'information visuelle cherchée devra prendre en compte la complexité des affichages où s'effectue la recherche. Pour des scènes peu complexes (hétérogénéité visuelle et faible niveau de détail), la présence de messages oraux d'assistance au repérage des cibles sera d'un faible intérêt. Elle pourra même s'avérer inutile et contribuer à alourdir inutilement la charge cognitive de l'utilisateur. Dans de tels contextes, on préférera une mise en relief visuelle. En revanche, le recours à une aide orale facilitera considérablement la recherche visuelle dans des scènes complexes (homogénéité visuelle et niveau de détail élevé). Cette conclusion correspond à celle exprimée dans (Althoff *et al.*, 2001).

### 6.5 Paysages versus objets complexes

Pour évaluer l'influence du sujet des photographies, objet complexe ou paysage, sur les performances des sujets, nous avons regroupé les scènes dans deux ensembles (2880 observations par ensemble), selon qu'elles représentaient des objets complexes (ensemble O) ou des paysages (ensemble P). Les résultats sont présentés dans le tableau 7.

On observe des différences importantes entre les ensembles O et P, à la fois pour les temps moyens de sélection des cibles et pour les taux d'erreurs. Ces différences sont statistiquement hautement significatives pour les temps de sélection ($t=-7.57$ ; $p<.0001$) comme pour la précision de la sélection ($t=-3.04$ ; $p<.0001$).

| Sujet de la scène | Temps moyen (ms) | Ecart type (ms) | Taux d'erreurs % |
|---|---|---|---|
| Objet | 3235 | 4175 | 3.84 |
| Paysage | 4186 | 5297 | 5.53 |

**Tableau 7.** *Temps moyens de sélection des cibles et taux d'erreurs par thème (O, P). Les pourcentages ont été calculés sur le nombre total d'erreurs pendant la session. (2880 observations par thème, Objet ou Paysage)*

Ces résultats montrent que le repérage visuel, quel que soit le mode de présentation de la cible, est plus facile lorsque celle-ci est un objet complexe plutôt qu'un paysage. Dans la condition PV (voir 6.1), il semble que la plupart des sujets, pendant la présentation visuelle de la cible, tentent de caractériser celle-ci verbalement pour faciliter sa détection ultérieure. Le manque d'indices visuels verbalisables facilement dans les photographies de paysages, notamment celles utilisées pour construire les scènes de difficulté 2 et 3, homogènes visuellement, rend cette stratégie inefficace. A niveau de difficulté égal, les paysages présentent en général un niveau de détail plus élevé que les objets, et une saillance visuelle moindre pour les paramètres forme et direction ; ils sont perçus davantage comme des textures plutôt qu'en termes de contrastes et d'opposition forme(s)/fond Cette interprétation semble compatible avec le modèle temporel « coarse to fine » de perception des fréquences spatiales (Huges *et al.*, 1996).

## 7 Conclusions générales

L'objectif de l'étude présentée était d'évaluer l'utilité effective de la multimodalité parole + graphique en tant qu'expression complémentaire du système dans des situation d'interaction mettant en jeu des activités de recherche visuelle dans des affichages complexes.

Nous avons adopté une démarche expérimentale pour étudier l'influence d'indications spatiales formulées oralement par le système sur les performances et la satisfaction d'utilisateurs potentiels ou, plus précisément, pour déterminer l'assistance effective fournie par de telles indications à des activités d'exploration visuelle d'affichages 2D présentant des collections de photographies (30 photographies par affichage). La tâche visuelle choisie, le repérage d'une cible familière visuellement (une photographie au cas particulier), intervient dans de nombreuses activités interactives, comme la recherche dans les visualisations de masses de données, la navigation sur le Web ou l'inspection visuelle.

Cette étude a montré que la présentation multimodale de la cible, à savoir, sa présentation visuelle accompagnée d'une indication spatiale absolue sur sa position dans la scène, est plus efficace que sa seule présentation visuelle, en termes de temps de recherche et de précision de la détection. Ce résultat incite à privilégier l'assistance orale au repérage lorsque l'affichage est suffisamment complexe pour qu'une mise en relief visuelle de la cible risque de passer inaperçue et/ou d'aggraver la surcharge visuelle de l'utilisateur.

La comparaison des performances des sujets en fonction du mode de présentation de la cible, l'étude qualitative de l'évolution de leurs performances au cours de la réalisation des 120 tâches proposées dans chaque condition, ainsi que l'analyse des questionnaires et débriefings post-expérimentaux, ont mis en évidence les bénéfices que les utilisateurs peuvent retirer, dans des situations d'interaction appropriées, de messages oraux d'assistance au repérage de cibles connues visuellement, non seulement en termes de précision et de rapidité, mais aussi de confort visuel. Des messages contenant de brèves indications spatiales absolues, indépendantes de l'organisation spatiale de l'affichage, épargnent à l'utilisateur, en restreignant la zone de recherche, une exploration exhaustive de la scène, source de fatigue oculaire et cognitive lorsque les recherches se multiplient.

Nous avons montré, en outre, que l'apport de messages oraux contenant des informations spatiales destinées à faciliter la recherche visuelle variait en fonction du niveau de difficulté de la scène, et donc, de celui de la tâche. L'assistance orale est plus efficace pour des scènes dont les items sont similaires visuellement, donc dénués de saillance visuelle et faciles à confondre les uns avec les autres surtout si le niveau de détail de chaque item est élevé. L'apport des messages multimodaux se révèle moindre, en revanche, sur des scènes dont les constituants sont hétérogènes visuellement, comportent peu de détails et sont familiers à l'utilisateur.

Cette étude a également permis de vérifier l'hypothèse selon laquelle l'organisation spatiale des items dans l'affichage influence l'efficacité et le confort de la recherche visuelle. En particulier, la structure radiale paraît mieux adaptée que les autres structures considérées aux informations spatiales choisies, pour assister la recherche visuelle. La mise en évidence de l'influence de la structure spatiale des affichages sur l'efficacité et le confort de la recherche visuelle nous incite à approfondir l'étude de cette influence en tentant, d'une part, d'expliquer sa relation avec les parcours oculaires spontanés des utilisateurs, ce qui implique la collecte de données oculométriques, et d'autre part, de déterminer dans quelle mesure cette influence persiste lorsqu'on augmente le nombre des items affichés, par exemple en le portant à 100, ce qui permettrait d'obtenir des résultats directement applicables à la navigation dans les visualisations de grands ensembles d'informations multimédias.

Enfin, les choix de conception du protocole expérimental se sont avérés pertinents, concernant le sujet des photographies (objets versus paysages), le niveau de difficulté des scènes (facile, moyen, difficile) et la structure spatiale des affichages (aléatoire, elliptique, radiale et matricielle), dans la mesure où nous avons observé, pour chacune de ces variables, des différences statistiquement significatives. Ces résultats impliquent la nécessité de tenir compte de ces facteurs d'influence (sujet/thème des scènes, difficulté de l'activité de recherche visuelle, organisation spatiale de l'affichage) lors de l'étude expérimentale de tâches visuelles et dans la conception des affichages graphiques des applications interactives.

## Bibliographie

**Annexes**

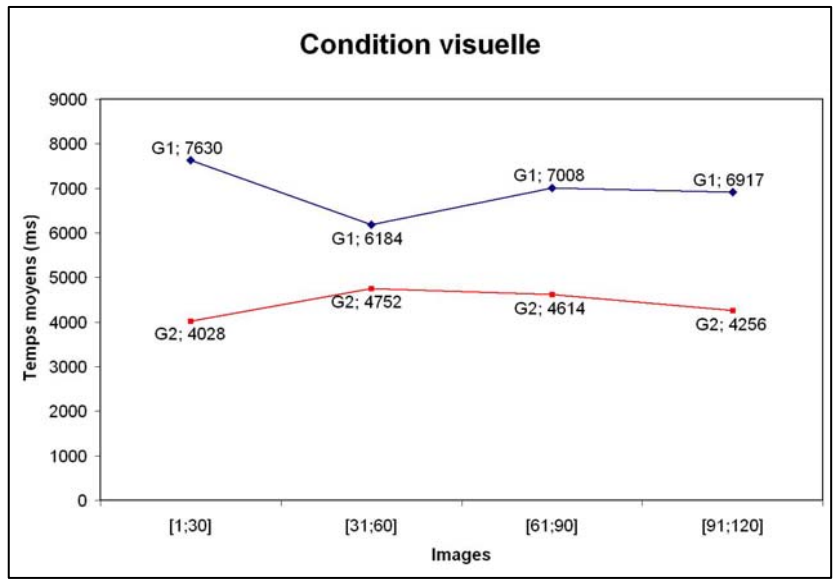

**A1.** *Évolution des temps moyens de sélection des cibles (condition PV)*
*G1 : ordre de passation PV-PM ; G2 : ordre de passation PM-PV*

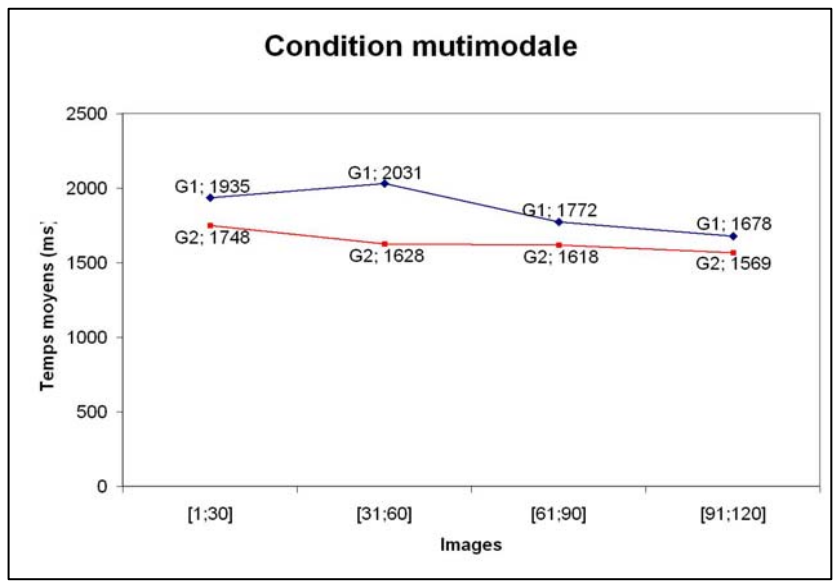

**A2.** *Évolution des temps moyens de sélection des cibles (condition PM)*
*G1 : ordre de passation PV-PM ; G2 : ordre de passation PM-PV*

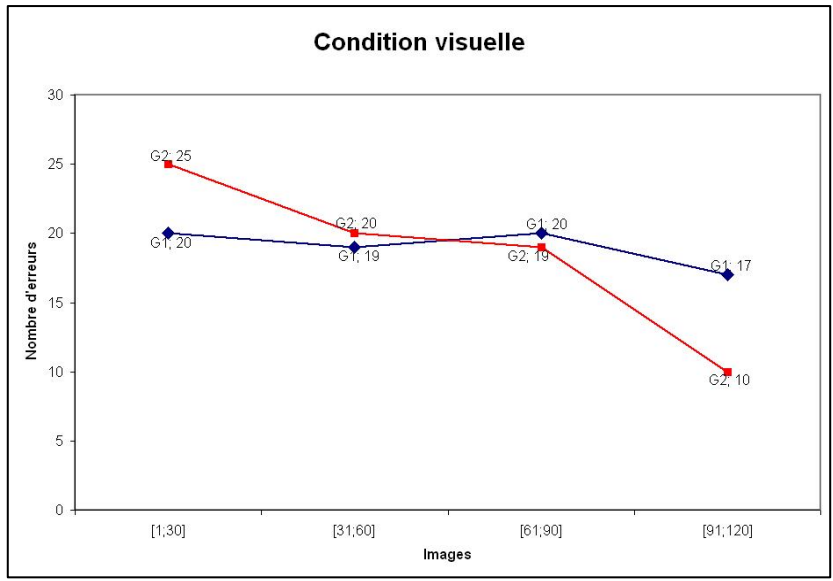

**A3.** *Évolution de la précision des sélections des cibles (condition PV)*
*G1 : ordre de passation PV-PM ; G2 : ordre de passation PM-PV*

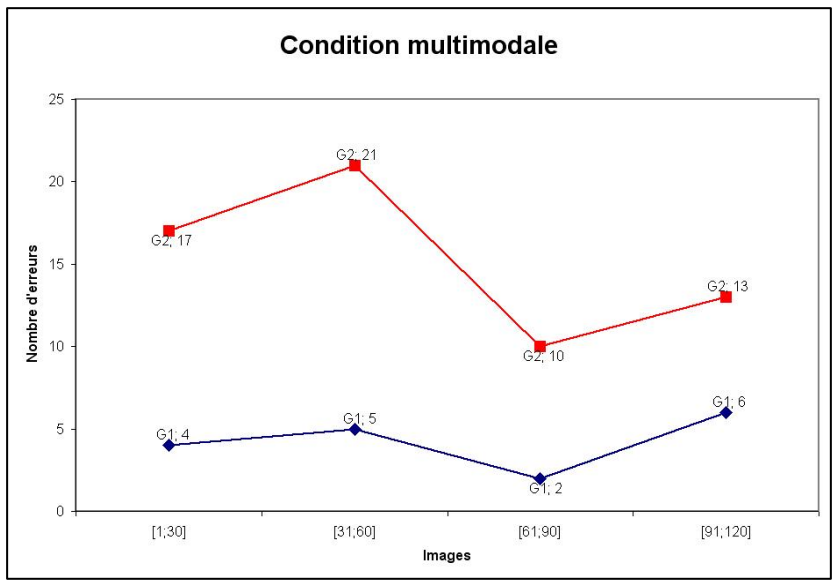

**A4.** *Évolution de la précision des sélections des cibles (condition PM)*
*G1 : ordre de passation PV-PM ; G2 : ordre de passation PM-PV*